\documentclass[conference]{IEEEtran}
\ifCLASSINFOpdf
\else
\fi
\usepackage{multirow}
\usepackage{cite}
\usepackage{amsmath,amssymb,amsfonts}
\usepackage{graphicx}
\usepackage{textcomp}
\usepackage{xcolor}
\usepackage{tabularx}
\usepackage{makecell}
\def\BibTeX{{\rm B\kern-.05em{\sc i\kern-.025em b}\kern-.08em
    T\kern-.1667em\lower.7ex\hbox{E}\kern-.125emX}}

\usepackage{booktabs}
\usepackage{algpseudocode}
\usepackage{algorithm}
\usepackage{amsmath}
\usepackage{amssymb}
\usepackage{multirow}
\usepackage{subcaption}
\usepackage{ulem} % for \dotuline
\usepackage{tikz} % for more complex underlines if necessary

\usepackage{fancyhdr}
\usepackage{lipsum}  % This package provides filler text

\usepackage[bookmarks=false]{hyperref}

\begin{document}
\title{MoodCam: Mood Prediction Through Smartphone-Based Facial Affect Analysis in Real-World Settings}

% \author{
% \IEEEauthorblockN{
% Rahul Islam \href{https://orcid.org/0000-0003-3601-0078}{\includegraphics[scale=0.06]{Fig/orcid.png}},
% Tongze Zhang \href{https://orcid.org/0000-0002-3375-7136}{\includegraphics[scale=0.06]{Fig/orcid.png}},
% Priyanshu Singh Bisen \href{https://orcid.org/0009-0007-5433-0020}{\includegraphics[scale=0.06]{Fig/orcid.png}},
% Sang Won Bae \href{https://orcid.org/0000-0002-2047-1358}{\includegraphics[scale=0.06]{Fig/orcid.png}}
% }
% \IEEEauthorblockA{Charles V. Schaefer, Jr. School of Engineering and Science\\
% Stevens Institute of Technology\\
% Hoboken, USA\\
% % mislam5@stevens.edu, sbae4@stevens.edu
% }
% }


\author{
\IEEEauthorblockN{
Rahul Islam \href{https://orcid.org/0000-0003-3601-0078}{\includegraphics[scale=0.06]{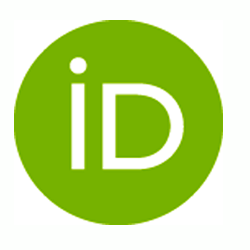}},
Tongze Zhang \href{https://orcid.org/0000-0002-3375-7136}{\includegraphics[scale=0.06]{Figs/orcid.png}},
Sang Won Bae \href{https://orcid.org/0000-0002-2047-1358}{\includegraphics[scale=0.06]{Figs/orcid.png}}
}
\IEEEauthorblockA{Charles V. Schaefer, Jr. School of Engineering and Science\\
Stevens Institute of Technology\\
Hoboken, USA\\
% mislam5@stevens.edu, sbae4@stevens.edu
}
}


\maketitle
% \thispagestyle{firstpage}

\begin{abstract}
MoodCam introduces a novel method for assessing mood by utilizing facial affect analysis through the front-facing camera of smartphones during everyday activities. We collected facial behavior primitives during 15,995 real-world phone interactions involving 25 participants over four weeks. We developed three models for timely intervention: momentary, daily average, and next day average. Notably, our models exhibit AUC scores ranging from 0.58 to 0.64 for Valence and 0.60 to 0.63 for Arousal. These scores are comparable to or better than those from some previous studies. This predictive ability suggests that MoodCam can effectively forecast mood trends, providing valuable insights for timely interventions and resource planning in mental health management. The results are promising as they demonstrate the viability of using real-time and predictive mood analysis to aid in mental health interventions and potentially offer preemptive support during critical periods identified through mood trend shifts.
\end{abstract}

\begin{IEEEkeywords}
Affective computing, Mood Detection, Machine Learning, Mobile computing, System, Empirical study that tells us about people, Application Instrumentation, Field Study
\end{IEEEkeywords}

% \fi
\IEEEpeerreviewmaketitle

\section{Introduction}
In today's world, smartphones are integral to everyday life, offering a unique opportunity to enhance user experience and monitor mental health by understanding users' moods through their interactions with these devices. With built-in feature such as face recognition systems, users frequently unlock their phones, enabling the front-facing camera to capture their natural expressions multiple times daily. These candid images, unlike posed selfies, are free from social influences and accurately reflect our emotional states. Our work expands upon recent initiatives like MoodCapture \cite{nepal2024moodcapture}, FacePsy \cite{islam2024facepsy}, PupilSense \cite{islam2024pupilsense}, and MoodPupilar \cite{islam2024moodpupilar} for mental health sensing, such as depression detection using user images collected from smartphones. This contributes towards a future where AI, integrated into our phones, can analyze images in real time using advanced machine learning techniques, determining our mood directly on the device to ensure privacy. This continuous, seamless method of mood monitoring signifies a shift in mental health assessment through facial behavior analysis, paving the way for instant mood identification, timely intervention, and ongoing monitoring. It can enhance interactions with smart devices like therapy chatbots, setting the stage for significant advancements in mental health care.

Mood is an affective state that plays a key role in our mental well-being and overall health. Yet, in the realm of mobile mental health care, tools that intuitively understand and respond to our emotional states without explicit input are notably absent. Although our devices can capture the physical environment through various sensors, they lack the ability to perceive the user's mood. However, the vast data generated from daily smartphone interactions, including pictures, text communications, app usage patterns, and social media activity, provides a valuable and realistic insight into a user's mood. The existing approach to mood tracking primarily depends on self-reported surveys, which are usually recorded as a mood diary for the user’s reference. These mood diary could be found in Apple Fitness or Google Fit, which rely on the user's subjective and potentially biased recall of their mood at any given time. Most recently several studies have utilized this smartphone sensing data to assess mood \cite{likamwa2013moodscope, mehrotra2017mytraces, kang2023k, meegahapola2023generalization, jacob2023affect}. This development allows virtual agents, chatbots, and health applications to provide more personalized and empathetic care by recognizing users' emotional states. It's a significant step towards making mental health support more integrated, accessible, and responsive through everyday digital use.

Most prior studies using facial affects data have been conducted in lab-controlled settings \cite{skaramagkas2023esee, sarkar2023avcaffe}, or utilize datasets curated from videos and media sharing platforms \cite{kossaifi2017afew, mollahosseini2017affectnet, kollias2021analysing, kollias2023abaw}. These datasets are typically used for challenges. The facial expressions captured in these datasets may not be genuine, as they are often performative and could be affected by biases such as social desirability. Furthermore, the generalizability of these studies is often limited due to their controlled environments and lack of real-world context. Therefore, more ecologically valid studies are needed to utilize data derived from naturalistic settings. This underscores the importance of developing methodologies to collect and analyze data in naturalistic settings, which can be more representative of real-world experiences and behaviors. This paper proposes a solution based on the advancements in smartphone cameras. We introduce MoodCam, a system designed to capture facial affect and self-reported mood in real-world settings using a smartphone. The facial data obtained capture authentic and spontaneous expressions, reducing the impact of self-consciousness on emotional expressions and increasing the data reliability. 

Our contributions are as follows:

\begin{enumerate}
  \item We have propose MoodCam, a passive mood-sensing system that automatically gathers facial affect data from users' front-facing smartphone cameras in real-world settings. This method ensures unobtrusive data collection. Our findings demonstrate that it's feasible to infer mood at any given moment, as well as daily and next day averages. The predictive abilities of MoodCam suggest it can effectively project mood trends, providing critical insights for timely interventions and resource planning in mental health management.
  
  \item We discuss the implications and application of MoodCam models. This system integrates into the existing ecosystem for the dissemination of mental health resources.
  
\end{enumerate}

\section{Related Work} \label{sec:relatedWork}
\subsection{Understanding Mood and Evaluation Metrics}
Mood represents a pervasive and sustained affective state that significantly influences an individual's perception and behavior. It has been extensively studied in psychology \cite{batson13distinguishing, beedie2005distinctions, likamwa2013moodscope}. Unlike emotions, which are typically intense and fleeting reactions to specific stimuli, moods are less intense but more enduring, often lasting several hours or even days \cite{beedie2005distinctions}. Moods differ from emotions in their longevity and in being less tied to specific events; they arise from a series of events and linger, subtly coloring one's psychological landscape \cite{beedie2005distinctions}. Mood can be understood as a background emotional tone that influences an individual's ongoing experience and behavior. While emotions are immediate and reactive, often observable by others, moods are more internalized and can govern one’s feelings over an extended period. This prolonged nature makes moods less visible but crucial in shaping one’s overall affective state \cite{beedie2005distinctions}.

A significant theoretical approach to understanding mood is the circumplex model of affect (CMA), initially proposed by Russell in 1980 \cite{russell1980circumplex}. This model posits that all affective experiences are derived from two fundamental neurophysiological systems: valence, ranging from pleasure to displeasure, and arousal, which varies from low to high alertness \cite{posner2005circumplex}. According to this framework, each emotional state is essentially a combination of these two dimensions. For instance, joy is viewed as a result of high positive valence and moderate arousal, reflecting a specific activation pattern within these systems \cite{posner2005circumplex}. Similarly, other affective states emerge from different activation levels across these dimensions, shaped further by how we cognitively interpret and label these underlying physiological sensations \cite{russell1980circumplex}. Mood is a significant and persistent affective state that colors an individual's perception and interaction with the world, distinct from the more transient and specific reactions characterized as emotions. The Circumplex model of affect provides a clear and practical framework for understanding and measuring these mood states, enhancing our ability to study and interact with the broad spectrum of human affects \cite{russell1980circumplex}.

Another prevalent method in the study of affect involves categorizing emotions into discrete, basic categories, as Ekman \cite{ekman1971universals, tomkins1962affect} popularized, identifying six primary emotions: happiness, sadness, fear, anger, disgust, and surprise \cite{ekman1971universals}. This model is intuitive but may not capture the complexity of human emotions that can co-occur and overlap. Additional methods like the Positive and Negative Affect Schedule (PANAS) \cite{watson1988development, crawford2004positive} and the Self-Assessment Manikin (SAM) \cite{bradley1994measuring} provide nuanced assessments through checklists (20 items) and graphic scales, measuring a broad spectrum of emotional states in terms of positive and negative affect, and dimensions of pleasure, arousal, and dominance. However, their complexity can hinder frequent use in practical settings. In contrast, the Circumplex model of affect, preferred for its simplicity, reduces mood to two dimensions—pleasure and activeness—allowing for efficient and consistent classification of emotional states \cite{posner2005circumplex}, and has been widely adopted \cite{likamwa2013moodscope, mehrotra2017mytraces, kang2023k, meegahapola2023generalization, jacob2023affect}. This makes it ideal for extensive field studies requiring frequent mood assessments, highlighting its practicality for both academic research and real-world applications.

\subsection{Mood Inference and Detection Using Smartphones and Wearables in-the-Wild}
Recent trends in mood tracking technologies show a notable shift from traditional self-reported methods to advanced sensing approaches. While survey based approaches often face low compliance rates due to the burden of frequent user engagement \cite{schueller2021understanding}, self-report-based mood tracking, which employs tools like mobile phone-based mood charts and ecological momentary assessment (EMA) responses, has played a foundational role in promoting self-awareness and managing anxiety, particularly among clinically depressed populations \cite{chan2018review, meegahapola2020smartphone, bakker2018engagement, birney2016moodhacker}. As a response, researchers have conducted studies to leverage mobile phone sensors to create context-aware systems that infer mood with minimal user input. Studies that use sensors as proxies for mood, stress, and other personal attributes have shown promising results, with some achieving significant mood inference accuracies \cite{likamwa2013moodscope, meegahapola2020smartphone, lane2010survey, servia2017mobile, kang2023k, meegahapola2023generalization, jacob2023affect}. For instance, LiKamwa et al. \cite{likamwa2013moodscope} achieved a mood inference accuracy of up to 67\% by utilizing smartphone app usage data as a proxy for mood in a population model. Kang et al. \cite{kang2023k} demonstrated F1 scores of 0.543 for valence and 0.534 for arousal by combining data from smartphones and wearables. In a more recent large-scale study, Meegahapola et al. \cite{meegahapola2023generalization} developed a model to predict high and low valence, yielding an AUC of 0.51. Similarly, Jacob et al. \cite{jacob2023affect} utilized IMU data from smartphones to predict mood, achieving an F1 score of 0.897 for both valence and arousal. These developments reflect a growing trend towards passive, sensor-based mood tracking systems that reduce user burden while delivering accurate and timely mood assessments.

\begin{table*}[h]
\caption{\label{tab:releated_works} Studies on Predicting Mood Using Circumplex Model of Affect with Smartphone and Wearable in Real-World Settings}
\centering  
\footnotesize
\begin{tabular}{p{3.5cm}p{.5cm}p{1.5cm}p{2.5cm}p{3cm}p{3cm}}
\toprule
\textbf{Related Work} & \textbf{Part.}  & \textbf{Study Length} & \textbf{Predictive Window} & \multicolumn{2}{c}{\textbf{Performance Metrics}} \\
 \cmidrule(lr){5-6}

 &  & & &  \textbf{Valence} & \textbf{Arousal} \\
\midrule
MoodScope\cite{likamwa2013moodscope} (2013) & 32 & 2 months  & Daily Mood  & Acc: 66\% & Acc: 67\%  \\

MyTraces\cite{mehrotra2017mytraces} (2017) & 28 & 6 months  & - & - & -  \\

K-EmoPhone\cite{kang2023k} (2023) & 77 & 7 days  & At moment & F1: 0.543 & F1: 0.534  \\

Meegahapola et al.\cite{meegahapola2023generalization} (2023) & 678 & 13 weeks & At moment & AUC: 0.51 & -  \\

Jacob et al. \cite{jacob2023affect} (2023) & 52 & 2 months & At moment & F1: 0.897 & F1: 0.897  \\

\textbf{*Our MoodCam Study} & 25 & 4 weeks &  At moment & F1: 0.55, AUC: 0.64 & F1: 0.72, AUC: 0.63  \\
&  & &  \textbf{*Daily Mood } & \textbf{F1: 0.77, AUC: 0.64} & \textbf{F1: 0.66, AUC: 0.58}  \\
&  & &  \textbf{*Next Day Mood} & \textbf{F1: 0.85, AUC: 0.61} & \textbf{F1:0.75, AUC: 0.63}  \\
\bottomrule
\end{tabular}
\end{table*}

Unlike passive sensors like GPS or activity tracking, which infer mood based on movement patterns or locations, facial expressions provide a direct, visual representation of an individual's emotional state. Recent systems like MoodCapture \cite{nepal2024moodcapture} and PupilSense \cite{islam2024pupilsense} utilizes spontaneous facial and eye images captured by the front-facing camera on smartphones, which helps in detecting unguarded emotional expressions for detecting depression. This is seen as more reliable for assessing mental states because it's not influenced by social desirability or self-presentation biases. Most people use smartphones daily and frequently unlock their phones using facial recognition. This regular interaction allows the capture of facial images without requiring additional actions or behavioral changes from users, thus ensuring the approach is unobtrusive and fits seamlessly into daily routines \cite{nepal2024moodcapture}. Facial recognition technology can function continuously as part of the smartphone's standard operation, providing real-time updates and ongoing monitoring of the user's emotional state. This is particularly valuable for mental health tracking, where changes can occur rapidly and unexpectedly. However, using facial affect data such as Action Units, Head Pose and others in affect modeling is not a new concept. Evaluating mental health and emotions through facial feature extraction has been a significant focus in computer vision. This has applications in various sectors including education \cite{islam2023microflow} and behavior modeling domain \cite{mellouk2020facial, song2020spectral, valstar2014avec, cohn2009detecting}. Previous studies have been conducted in lab-controlled settings \cite{skaramagkas2023esee, sarkar2023avcaffe}, or used datasets curated from videos and media sharing platforms \cite{kossaifi2017afew, mollahosseini2017affectnet, kollias2021analysing, kollias2023abaw} for predicting valence and arousal. Our work aims to extend these studies by bridging the gap between the utility of facial affect data in lab-controlled studies and real-world contexts.

% \textbf{Comparison to MoodCam.} \textcolor{red}{Ask Grace for comment on comparision}
\section{User Study and Field Data Collection\protect\footnote{Please see our paper \cite{islam2024facepsy} for more details.}}

\subsection{Research Design and Participant Demographics}
\subsubsection{Protocol}
The study received approval from the Institutional Review Board (IRB) at the University and was conducted remotely using Zoom during the COVID-19 Public Health Emergency (PHE). Participants, who were required to be at least 18 years old and own an Android phone with a data plan, joined the study from various time zones. They completed a preliminary screening survey and scheduled their onboarding video call. During this session, the interviewer provided informed consent information and administered baseline questions. The study's mobile app, which was used to capture facial behavior primitives (See Section \ref{sec:facialBehaviorDataLogger}), was then installed on the participants' phones. Study questionnaires were sent and completed through notifications using the Qualtrics online survey platform. Participants were instructed to complete daily mood rating surveys three times a day in order to receive full compliance. We suggested, but did not require, that they should complete these tasks around 10am, 3pm, and 8pm in their local time zone. Participants were eligible to receive up to \$135 for full participation in the study. They were compensated \$20 for completing the baseline assessment and installing the data collection app, and an additional \$25 per week for a duration of four weeks.

\subsubsection{Participants}
We recruited 38 participants for our user study, of which only 25 participants completed it. At the start of the study, one participant withdrew after two days, citing the app for excessive battery usage. It was later found that the high battery drain was due to their heavy social media usage, which often triggered the data collection process. Additionally, three participants left due to personal reasons, five did not complete the necessary surveys, and four were excluded because their Android devices had versions that were not compatible with our data collection trigger module. Therefore our analysis consisted of 11 male and 8 female participants, along with 6 unspecified entries, bringing the total to 25 participants. The average age of participants was 27.88 years, with males averaging 24.11 years and females 32.71 years. In terms of ethnicity, the participants primarily identified as Asian (15) or Caucasian (4), with 6 unspecified. Educational backgrounds varied, including 1 participant with a high school diploma, 8 with a bachelor's degree, and 10 with a master's degree. Mental health was rated on average at 6.79 out of 10, slightly higher among females (6.88) compared to males (6.73). Regarding mental health specifics, our data shows that depression was experienced 'not so often' by 8 participants and 'somewhat often' by 6. A significant finding was that 2 participants, both females, reported experiencing depression 'very often.' Additionally, 4 participants confirmed having a diagnosed mental disorder. The study also looked at marijuana use, where a majority (17) reported not using marijuana, and only 2 participants, both male, confirmed usage. 

\begin{figure*}[h]
    \centering
    \includegraphics[width=1\textwidth]{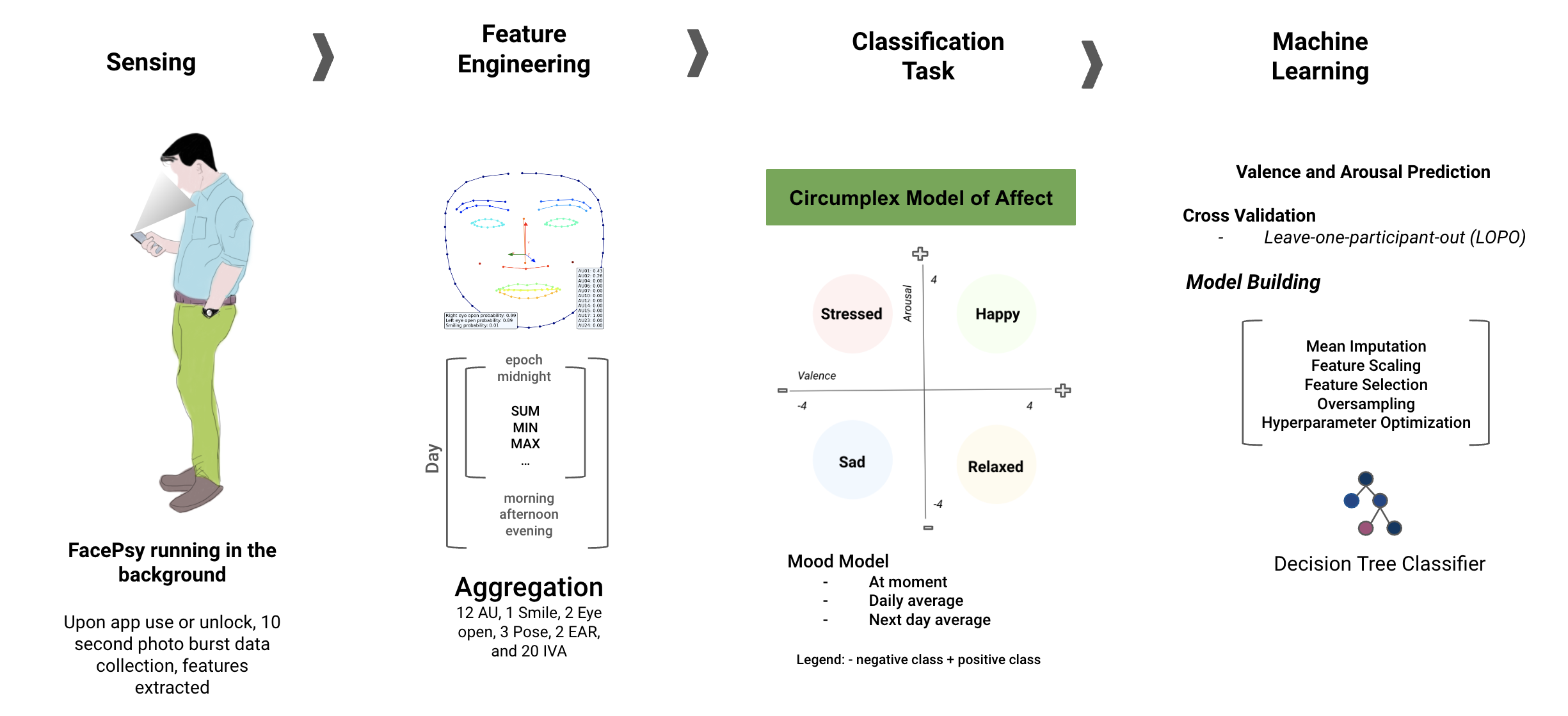}
    \caption{Proposed MoodCam system and models, analyzing facial affect data from real-world settings}
    \label{fig:cma}
\end{figure*}

\subsection{Field Study}

\subsubsection{Data Collection} \label{sec:facialBehaviorDataLogger}
In our research data collection, we employed the affective mobile sensing system, FacePsy \cite{islam2024facepsy}, which runs in the background on Android smartphones. This system opportunistically captures images of the facial behavior primitives whenever the user unlocks their phone or launches any predefined trigger apps, which are categorized into communication, social media, productivity, entertainment, and health. Upon activation, the data collection session is initiated for a 10-second photo burst; the system logs a 151-channel time-series dataset for each data collection session, encompassing 12 AU intensities, 1 smile probability, 2 probabilities related to eyes being open, 3 head pose metrics, and 133 facial landmarks as frame-by-frame descriptors. The app also includes a feature to collect daily mood data through notifications. It sends notifications to complete daily mood rating (i.e., CMA) surveys three times a day. These notifications are scheduled to appear at 10am, 3pm, and 8pm.
For detailed information on FacePsy implementation and use please see their paper \cite{islam2024facepsy}.

\subsubsection{Dataset}
The application gathered facial behavior data from participants, totaling 15,995 instances (equivalent to data collection triggers), with an average of 639.8 phone interactions per participant for data collection. Over a four-week period, we amassed 544 days of facial data from 25 participants. However, there were variations in how long each participant stayed in the study; on average, participants missed 3.36 days, resulting in a cumulative 616 participant days. Notably, 55 of these days recorded no data due to reasons such as participant-reported vacations or breaks. Additionally, 17 days of data were unusable due to issues like poor image quality and low eye-open probability during feature extraction. As a result, the total effective data available for analysis was 544 days.

The participants were sent notifications three times daily for 4 weeks at 10 am, 3 pm, and 8 pm to report their mood administered with CMA. We suggested that participants report their moods at each instance. We collected 577 days of CMA survey from our participants; on average, we found that participants reported 2.23 daily mood reports. We then map these ground truth values in the facial behavior dataset.

\section{The Design of MoodCam Models}
The key feature of MoodCam is its ability to assess a user's mood at three different times: the present moment, the daily average, and the following day. Building on the data and findings from Section \ref{sec:relatedWork}, this section explores the development of MoodCam's predictive models. We use various supervised machine learning methods to examine the possibility of predicting a user's mood through facial behavior analysis. We discuss the generation of mood samples for immediate, daily, and next day assessments, as well as the derivation of facial behavior features. Finally, we explore the creation and testing of different mood prediction models based on our mood surveys and facial behavior data.

\subsection{MoodCam Output}
The CMA survey, which was administered to participants, assesses mood using two scales: valence and arousal, each ranging from -4 to 4. We learned that majority of the previous
works in this area modeled valence and arousal in binary classification problems \cite{likamwa2013moodscope, meegahapola2023generalization, wampfler2022affective}. Therefore for modeling purposes, these scales have been converted into a binary classification problem. Specifically, scores less than 0 are categorized as low valence or arousal, while scores of 0 or higher are classified as high valence or arousal. This binary classification strategy is consistently applied in the construction of mood models for both valence and arousal, as discussed in the subsequent section below.

\subsubsection{At moment}
We estimate the user's mood at any given moment. To achieve this, we use a 30-minute time window for predicting mood. This approach is based on prior research \cite{yang2021empirical}, which suggests a time-sensitive relationship between context and activity. Other research \cite{schwerdtfeger2010momentary} has found that momentary affect predicts bodily movement in the subsequent 30 minutes. Moreover, a study \cite{giesbrecht2012psychological} implemented a 30-minute window for mood assessment to account for possible delays between the onset of a stressor and observable changes in cortisol levels, as well as the internal mood experience. Taking into consideration this, we decided to label any facial behavior data collected within 30 min prior to when participant responded to mood to their respective mood survey data. If the user's mood could be predicted at moment, effective and timely interventions could be delivered during risky moments such as periods of stress or sadness.

\subsubsection{Daily Average}
Prior studies \cite{likamwa2013moodscope} in mood sensing have worked on the output of their model as daily averages by taking the mean of all the mood samples collected throughout the day. Consequently, we also build a model that predicts daily averages. To develop the model, we divided each participant's day into four 6-hour segments: midnight, morning, afternoon, and evening. We calculated statistical features like the min, max, mean, median, sum, std, q1 and q3 for each epoch based on the facial behavior data. We then label them to respective daily average mood. If daily mood averages can be reliably predicted, they can be utilized for several purposes. Beyond just reflecting a user's current emotional state, observing trends in these daily mood averages can signal when a user's mood is deviating from their usual pattern.

\subsubsection{Next Day Average}
Another useful model predicts future moods based on past days' data. This model can provide a mood forecast, allowing for adequate preparation of resources, such as therapists. This is particularly beneficial when resources are scarce or when there's a need to closely monitor mood changes. To create this model, we split each participant's day into four 6-hour segments: midnight, morning, afternoon, and evening. We used facial behavior data to calculate statistical features such as the min, max, mean, median, sum, std, q1 and q3 for each segment. We tested different lags (1, 2, 4, and 8 previous days of data to predict the next day) to predict the mood for the next day's average.

\subsection{Facial Behavior Primitives as MoodCam Input\protect\footnote{Please see our paper \cite{islam2024facepsy} for more details.}}
The FacePsy system processes most features directly on the user's phone, minimizing the need for extensive post-processing. It pre-extracts a range of features, including specific Action Units (AU1, 2, 4, 6-7, 10, 12, 14, 17, 23-24), as well as probabilities for smiling and eyes being open, facial landmarks, and head pose metrics like yaw, pitch, and roll. Additionally, the system captures other on-device features such as the Eye-aspect ratio and Inter-vector angles. Further details on these features are provided below.

\subsubsection{Inter-vector angle}
Inter-Vector Angles (IVA) are scale-invariant geometric features extracted from facial landmarks to represent facial shapes, as described in \cite{islam2016sention}. The facial centroid, located at the center of the nose, is used to calculate these IVA features. The face is segmented into eight regions: the nose center, jawline, left eyebrow, left eye, right eyebrow, right eye, mouth, and cheeks. A total of 1439 triangles are formed by generating all possible triangle permutations that extend from the centroid to other facial landmarks. To handle the complexity and large dataset, Principal Component Analysis is applied to reduce the IVA features to 10 principal components. Additionally, angular velocity and acceleration are computed.

\subsubsection{Eye-aspect ratio}
The Eye Aspect Ratio (EAR) is a metric utilized to assess the openness of the eye. It's calculated by dividing the sum of the eye's vertical dimensions by twice its horizontal length.

In total, we collected 12 Action Units (AU), 1 Smile Probability, 2 Eye Open Probabilities, 3 Head Euler Angles, 2 Eye Aspect Ratios, and 20 Inter-Vector Angles.

\subsection{Internal Model of MoodCam}
We use classification algorithms to develop inference models for our at-the-moment, daily, and next-day mood labels, using facial behavior data as input features. To identify the most relevant features and reduce the complexity of our data, we employ feature selection techniques using a random forest algorithm. We assess the effectiveness of our models through a leave-one-participant-out (LOPO) cross-validation method, also known as leave-one-out or leave-one-group-out. This technique is widely utilized in mobile inference systems and helps gauge the rigorous generalizability of our models. Once established, this classifier is fixed and not subject to changes.

\subsubsection{Feature Selection}
We utilized a Random Forest classifier for feature selection, which assessed the importance of each feature within our dataset. Based on the mean of the importance scores (Gini importance) generated by the classifier, we established a selection criterion. Features with importance scores above the mean were deemed essential and retained for further analysis, while those scoring below the mean were excluded. This data-driven, unbiased approach enhances model interpretability and performance by concentrating on features most critical for predicting the user's mood.

% \subsubsection{Regression}

\subsubsection{Classification}
Our classification system employs tree-based supervised machine-learning algorithms, specifically Decision Trees. They are selected based on their demonstrated effectiveness in prior behavioral modeling studies \cite{yang2016decision, islam2018depression}. To address the sample imbalance between low/high valence and arousal in our dataset and to minimize bias, we used the synthetic minority over-sampling technique (SMOTE) \cite{chawla2002smote} on the training data. Furthermore, hyperparameter tuning was performed to optimize the area under the receiver operating characteristic curve (AUC). We assessed the effectiveness of our machine learning models using metrics like the F1 score and AUC. To ensure generalization across users, we created a standardized classifier for mood and validated it using the leave-one-participant-out (LOPO) cross-validation method.

\section{Results}

\subsection{Predictive Analysis}
Table \ref{tab:results_cma} presents the performance of four MoodCam models designed to predict mood at the moment, daily average, and next-day average with 1 and 2-day lags (i.e., previous day of data) for both valence and arousal. The 'At moment' model shows moderate predictive ability with an F1 score of 0.55 and an AUC of 0.64 for valence, and higher performance for arousal with an F1 score of 0.72 and an AUC of 0.63. The 'Daily Average' model demonstrates strong performance in predicting daily mood averages, crucial for monitoring shifts in a user’s typical mood pattern, with an F1 score of 0.77 and an AUC of 0.64 for valence. The 'Next Day Average' models, with 1-day and 2-day lags, perform robustly, particularly the 2-day lag model, which achieves the highest F1 score of 0.85 and an AUC of 0.61 for valence. The model performance stop improving after 2 days of lag. These predictive capabilities indicate that the MoodCam can effectively forecast mood trends, offering valuable insights for timely interventions and resource planning in mental health management. The implications of this performance are significant as they confirm the feasibility of using real-time and predictive mood analysis to aid in mental health interventions and to potentially provide preemptive support during critical periods identified through mood trend shifts.

\begin{table}[h]
\caption{\label{tab:results_cma}Results}
\centering  
% \footnotesize
\begin{tabular}{p{3.5cm}p{0.5cm}p{1cm}p{0.5cm}p{1cm}}
\toprule
\textbf{Mood Model} & \multicolumn{2}{c}{\textbf{Valence}} & \multicolumn{2}{c}{\textbf{Arousal}} \\
% \cmidrule(lr){2-3} \cmidrule(lr){4-5}
\cline{2-5} 
 & F1 & AUC  & F1 & AUC  \\
\hline
At moment & 0.55 & 0.64 & 0.72 & 0.63 \\ 
Daily Average & 0.77 & 0.64 & 0.66 & 0.58 \\ 
Next Day Average (1 Day Lag) & 0.78 & 0.60 & 0.75 & 0.63 \\ 
Next Day Average (2 Day Lag) & 0.85 & 0.61 & 0.67 & 0.60 \\ 
\bottomrule
\end{tabular}
\end{table}

\subsection{Ablation Study}
In the ablation study of the MoodCam models, we evaluated the impact of each individual feature set on the valence and arousal prediction accuracy. To do this, we selectively remove one feature at a time from the complete feature set and evaluate model performance in this subset of features. Our goal was to identify which features contribute the most to the model's predictive performance. Each feature type—Action Units, Eye Open probability, Smiling probability, Head Euler Angles, Eye Aspect Ratios, and Inter-Vector Angles—was evaluated independently to understand their utility in mood prediction. The models' performance metrics with each feature removed are detailed in Table \ref{tab:ablation}. The table shows results across four different mood prediction models: At moment, Daily, Next Day (1 Day Lag), and Next Day (2 Day Lag), measured by F1 score and AUC for both valence and arousal.

From Table \ref{tab:ablation}, several trends emerge that provide nature of contribution of each feature sets towards model performance. Notably, features computed from Eye Aspect Ratios show good performance across different prediction models, suggesting their strong predictive value, particularly in daily and next-day mood predictions. In contrast, the removal of Action Units generally leads to a decrease in AUC, especially in the arousal prediction for the 'At moment' model, indicating their importance in capturing immediate mood states.  Moreover, while features like Head Euler Angles show lower performance in certain models, their contribution varies, highlighting the importance of contextual application. Interestingly, the impact of removing Smiling and Eye Open probability on performance is mixed, suggesting their conditional utility in specific mood models.

\begin{table*}[h]
    \centering
    \footnotesize
    \caption{Ablation Study: Investigating mood prediction of MoodCam feature sets\label{tab:ablation}}
    \begin{tabular}{lcccccccccccccccc}
    \toprule
    \multirow{2}{*}{\textbf{Feature}} & \multicolumn{4}{c}{\textbf{At moment}} & \multicolumn{4}{c}{\textbf{Daily}} & \multicolumn{4}{c}{\textbf{Next Day (1 Day Lag)}} & \multicolumn{4}{c}{\textbf{Next Day (2 Day Lag)}} \\ 
    \cline{2-17} 
                                    & \multicolumn{2}{c}{Valence} & \multicolumn{2}{c}{Arousal} & \multicolumn{2}{c}{Valence} & \multicolumn{2}{c}{Arousal} & \multicolumn{2}{c}{Valence} & \multicolumn{2}{c}{Arousal} & \multicolumn{2}{c}{Valence} & \multicolumn{2}{c}{Arousal} \\ 
                                    \cline{2-17} 
                                    & F1 & AUC & F1 & AUC & F1 & AUC & F1 & AUC & F1 & AUC & F1 & AUC & F1 & AUC & F1 & AUC \\ 
                                    \hline
        Eye Open & 0.57  & 0.58  & 0.55  & 0.53  & 0.81  & 0.51  & 0.59  & 0.48  & 0.78  & 0.53  & 0.68  & 0.54  & 0.82  & 0.49  & 0.63  & 0.54  \\ 
        Smiling  & 0.51  & 0.56  & 0.57  & 0.51  & 0.83  & 0.65  & 0.61  & 0.50  & 0.77  & 0.58  & 0.72  & 0.52  & 0.81  & 0.53  & 0.66  & 0.50  \\ 
        Head Euler Angle & 0.50  & 0.53  & 0.45  & 0.46  & 0.77  & 0.53  & 0.65  & 0.46  & 0.77  & 0.47  & 0.66  & 0.49  & 0.81  & 0.50  & 0.57  & 0.38  \\ 
        Action Units & 0.47  & 0.56  & 0.28  & 0.62  & 0.79  & 0.45  & 0.64  & 0.51  & 0.77  & 0.50  & 0.72  & 0.57  & 0.82  & 0.58  & 0.65  & 0.54  \\ 
        Eye Aspect Ratios & 0.57  & 0.60  & 0.67  & 0.49  & 0.83  & 0.62  & 0.66  & 0.50  & 0.83  & 0.54  & 0.63  & 0.49  & 0.82  & 0.49  & 0.63  & 0.51  \\ 
        Inter-Vector Angles & 0.48  & 0.57  & 0.65  & 0.44  & 0.82  & 0.64  & 0.61  & 0.52  & 0.85  & 0.51  & 0.69  & 0.48  & 0.85  & 0.53  & 0.64  & 0.55  \\ 
    \bottomrule
    \end{tabular}
\end{table*}

In summary, this ablation study highlights the varying importance of different feature types in mood prediction models. By understanding these contributions, we can optimize MoodCam, enhancing its accuracy and reliability. We can tailor feature selection to each mood prediction model's specific needs. This insight helps refine the tool, enabling more precise and timely interventions—especially crucial for mental health applications where accurate mood tracking is vital.

\section{Discussion}
We study the potential of using facial affect analysis in a real-world setting for detecting the mood of a person into valence and arousal based on the circumplex model of affect, towards development of user-centered and unobtrusive mood assessment tools. Our analysis shows the performance of machine learning of such approaches. Our predictive analysis demonstrates that a decision tree model with feature selection with random forest trained on facial affect data on three timely instances: at moment, daily average and next day average. For the "At moment" model, the AUC scores are 0.64 for Valence and 0.63 for Arousal. The "Daily Average" model has consistent AUC scores for both Valence and Arousal at 0.64 and 0.58, respectively. The "Next Day Average (1 Day Lag)" model reports a slight decrease in Valence to 0.60, while Arousal remains consistent at 0.63. Lastly, the "Next Day Average (2 Day Lag)" model exhibits a small increase in Valence AUC to 0.61 and maintains Arousal at 0.60. These results indicate that the models' ability to predict mood varies, with the "At moment" and "Next Day Average (1 Day Lag)" models showing relatively higher AUC scores for Arousal compared to their performance in Valence. 

During our ablation study, we found certain features had better signals in mood evaluation. Specifically, we notice Eye Aspect Ratios consistently show high performance across different prediction models. In contrast, the removal of Action Units generally leads to a decrease in AUC, especially in the arousal prediction for the ’At moment’ model, indicating their importance in capturing immediate mood states. By focusing on these features, researchers/developers can reduce the load on feature processing, making it faster and resource-optimized models to detect mood on mobile devices.

Previous studies have explored facial affect data predicting valence and arousal in lab-controlled settings \cite{skaramagkas2023esee, sarkar2023avcaffe}, or using datasets compiled from videos and media sharing platforms \cite{kossaifi2017afew, mollahosseini2017affectnet, kollias2021analysing, kollias2023abaw}. However, the application of facial affect data for real-world mood assessment has not been extensively researched. Our study aims to address this gap. Real-world scenarios involve numerous uncontrollable variables, such as lighting and personal activities, which can introduce noise into the data and potentially obscure true emotional indicators. In contrast, controlled environments minimize external variables \cite{campbell2015experimental}, ensuring the emotional state of the subject primarily influences the data collected. In affective computing, human-computer interactions in real-world settings are complex and ever-changing. To design an effective mobile framework for mental health, systems must not only interpret emotional states but also account for the context in which interactions occur. Mood indicators impacted by context significantly, making them harder to interpret. Developing a framework should address the challenges of data collection in real-world settings, innovating on its implementation by incorporating adaptive algorithms that can adjust to different states, resulting in consistent and accurate predictions. To differentiate between true mood indicators and those influenced by external factors, integrating context-aware computing can enable the system for such use.

\textbf{Comparision to prior studies.} Our method demonstrates good performance in mood assessment compared to previous work by mobile sensing and wearables \cite{likamwa2013moodscope, kang2023k, meegahapola2023generalization, jacob2023affect}. However, it's important to note that this comparison may not be entirely direct. These studies use different sensing modalitiies and feature sets than in the present study. Notably, our models exhibit AUC scores that range from 0.58 to 0.64 for Valence and 0.60 to 0.63 for Arousal, which are generally in line with or better than some previous studies. For instance, Meegahapola et al. \cite{meegahapola2023generalization} reported an AUROC of 0.51 for Valence, which is lower than all our reported AUC scores. Similarly, K-EmoPhone \cite{kang2023k} reported F1 scores around 0.543 for Valence and 0.534 for Arousal, which are lower than most of our F1 scores, indicating our models might be better at capturing the complexities of mood variations. Additionally, Jacob et al. \cite{jacob2023affect} achieved very high F1 scores (0.897), significantly outperforming our models in terms of performance. However, considering the variety of metrics (F1 and AUC) and modalities used across these studies, our results are promising, especially given the challenging nature of predicting mood states withß real-world facial affect analysis. 

\subsection{Implications and Potential Applications of MoodCam Models}

Current methods for tracking mood include self-reported questionnaires, such as those found in Apple Fitness and Google Fit, as well as sensor-based techniques utilizing data from smartphones and wearables. Conventional approaches typically involve users recalling and documenting their mood, often serving as a personal mood diary. As an alternative, the MoodCam initiative delineates a sophisticated approach to mood assessment, employing three distinct models: "At Moment," "Daily Average," and "Next Day Average." Each model offers unique insights and serves different functions within the mental health care delivery ecosystem, facilitating timely and appropriate interventions.

The "At Moment" model operates on a 30-minute window, allowing for immediate mood predictions based on recent facial behavior data. This real-time capability is especially useful for integrating with digital interventions, such as therapy chatbots, which can provide instant support during critical moments when traditional human resources might not be readily available. This model supports the immediate deployment of digital mental health tools, acting as a first line of response in moments of acute stress or emotional distress, potentially stabilizing the situation until further help can be arranged.

On a broader temporal scale, the "Daily Average" and "Next Day Average" models provide deeper temporal insights, suitable for both short-term and predictive care planning. The "Daily Average" model, by observing mood fluctuations throughout the day, helps in identifying deviations from a user’s typical mood pattern. This can signal the need for interventions such as scheduled therapeutic sessions or proactive wellness checks, particularly useful in managing chronic mental health conditions. Meanwhile, the "Next Day Average" model forecasts future moods from past data, allowing mental health professionals to optimize resource allocation—scheduling therapist sessions or check-in calls for at-risk patients, thus ensuring that scarce resources like human therapists are utilized effectively and efficiently. Together, these models offer a comprehensive toolkit for modern mental health practitioners, enhancing existing care structures and enabling more dynamic, responsive, and personalized care delivery.

\section{Limitation}
Our study provides valuable insights into using real-world facial affect data for mood detection, but there are some limitations to consider. Although we developed a successful model for population-level mood detection, it might not capture individual-specific patterns, which could limit its general applicability. To address this, we plan to gather more extensive data for each participant in future research and develop personalized models. Currently, we track 11 categories of app usage, but adding more categories could enhance data collection. Future studies should investigate whether more categories would be helpful. Our study also relies on self-reported measures like CMA, which could be influenced by social desirability and recall errors. Incorporating objective mental health assessments, including clinical evaluations and physiological markers, could provide significant benefits. Another limitation is our study's primary focus on analyzing facial affect data in relation to depression. Factors like social interactions, app usage, physical activity, and lighting conditions could also be beneficial for mood detection. Incorporating these elements in future research could lead to the development of more comprehensive and accurate predictive models. We also plan to improve our model by incorporating contextual layers, such as categorizing apps that trigger data collection within the model development process. This enhancement, part of the data collection triggering mechanism, aims to better understand facial behavior in various contexts.
\section{Conclusion}
Through this study, we demonstrate the potential of using facial affect analysis from real-world scenarios and machine learning to detect mood, paving the way for unobtrusive mental health assessment. Previous studies were conducted in controlled lab settings or used datasets curated from videos and media sharing platforms for predicting emotional responses. Our work seeks to expand upon these studies by bridging the gap between the use of facial affect data in lab-controlled studies and in real-world contexts. In the MoodCam study, we gathered facial affect data from 15,995 real-world phone interactions. This involved 25 participants over a span of four weeks. We created three models for timely intervention, namely: momentary, daily average, and next day average. Our models notably exhibit AUC scores ranging from 0.58 to 0.64 for Valence and 0.60 to 0.63 for Arousal.


% \section*{Acknowledgment}




%\bibliographystyle{IEEEtran}


% \end{thebibliography}

\bibliographystyle{IEEEtran}
\bibliography{Reference}

\end{document}